\documentclass[11pt, oneside]{article}   	
\usepackage{geometry, float, amsmath, cases, amssymb, authblk}  
\usepackage{graphicx, hyphenat, subfig}				
\usepackage{xcolor}							
\usepackage[superscript,biblabel]{cite}				
\usepackage{color, soul}		
\graphicspath{ {Figures/}}

\newcommand{\dd}{\mathop{}\!\mathrm{d}}

\title{Accelerated vortex dynamics across the magnetic 3D-to-2D crossover in disordered superconductors}

\author[1, 2]{\small Serena Eley*}
\author[3]{\small Roland Willa}
\author[4]{\small Masashi Miura}
\author[4]{\small Michio Sato}
\author[1]{\small Maxime Leroux}
\author[5]{\small Michael David Henry}
\author[1]{\small Leonardo Civale}
\affil[1]{\footnotesize Condensed Matter \& Magnet Science, Los Alamos National Laboratory, Los Alamos, NM}
\affil[2]{\footnotesize Department of Physics, Colorado School of Mines, Golden, CO}
\affil[3]{\footnotesize Materials Science Division, Argonne National Laboratory, Argonne, IL}
\affil[4]{\footnotesize Graduate School of Science \& Technology, Seikei University, Tokyo, Japan}
\affil[5]{\footnotesize MESA Fabrication Facility, Sandia National Laboratories, Albuquerque, NM}

\begin{document}
\maketitle

\noindent Running Title: Vortex pinning length versus film thickness \\
Corresponding Author: Serena Eley \\
Address: Department of Physics, 1523 Illinois Street, Golden, Colorado 80401 \\
Telephone: 805-452-2457 \\
E-mail address: serenaeley@mines.edu \\

\noindent Funding Sources
\begin{itemize}
\item U.S. Department of Energy, Office of Basic Energy Sciences, Materials Science and Engineering Division
\item JSPS KAKENHI (17H03239 and 17K18888) and a research grant from the Japan Power Academy
\item LDRD program at Sandia National Laboratory (SNL)
\item Swiss National Science Foundation (SNSF) through the Early Postdoc.Mobility program.
\end{itemize}

\newpage
\section{Abstract}
Disorder can have remarkably disparate consequences in superconductors, driving superconductor\hyp{}insulator transitions in ultrathin films by localizing electron pairs\cite{Goldman1998} and boosting the supercurrent carrying capacity of thick films by localizing vortices\cite{Blatter1994b} (magnetic flux lines). Though the electronic 3D-to-2D crossover at material thicknesses $d \sim \xi$ (coherence length) is well studied, a similarly consequential magnetic crossover at $d \sim L_c$ (pinning length) that should drastically alter material properties remains largely underexamined.  According to collective pinning theory\cite{Larkin1979}, vortex segments of length $L_c$ bend to adjust to energy wells provided by point defects. Consequently, if $d$ truncates $L_c$, a change from elastic to rigid vortex dynamics should increase the rate of thermally activated vortex motion $S$. Here, we characterize the dependence of $S$ on sample thickness in Nb and cuprate films.  The results for Nb are consistent with collective pinning theory, whereas creep in the cuprate is strongly influenced by sparse large precipitates. We leverage the sensitivity of $S$ to $d$ to determine the generally unknown scale $L_c$, establishing a new route for extracting pinning lengths in heterogeneously disordered materials.

\section{Introduction}
\begin{figure}[t]
\centering
\includegraphics[width = .95\textwidth]{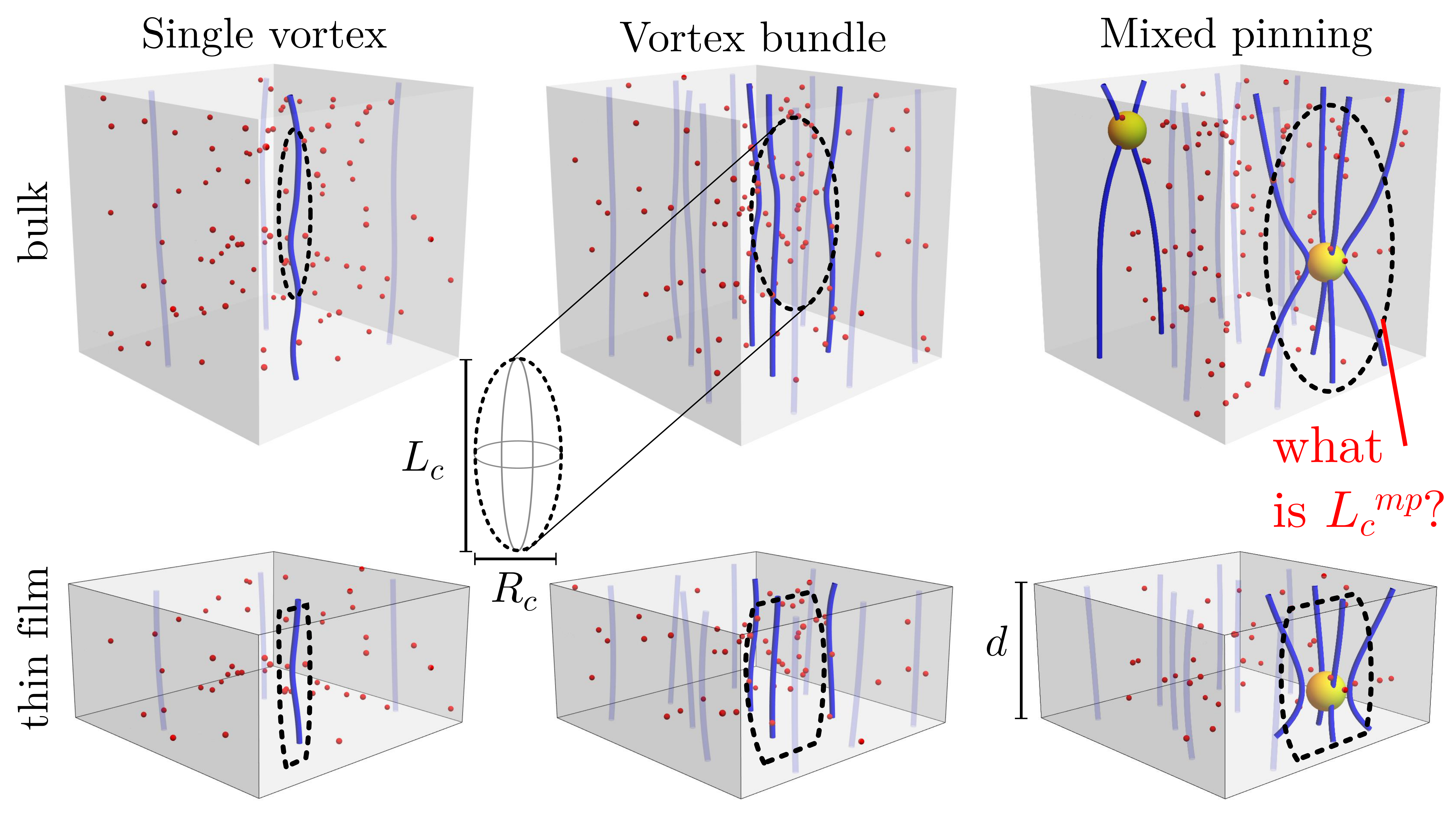}
\caption{\textbf{Correlation volumes in different pinning scenarios.} Illustration of pinning volumes for the single vortex, vortex bundle, and mixed pinning regime (from left to right). The volume $V_{c} = L_{c}R_{c}^{2}$ (indicated by a dashed ellipsoid) contains many point defects (red) and/or few large inclusions (yellow) and determines the creep rate. The bottom row indicates the situation where $L_{c}$ is truncated by the sample thickness $d$. The description of pinning in systems with heterogeneous defects poses an unsolved problem.
}
\label{fig:cartoon}
\end{figure}

Perfect crystallinity is an uncommon material characteristic.  For superconductors, it is not only rare, but also often undesirable, as defects are prerequisite for these materials to host high dissipationless currents.  This is because disorder can immobilize (pin) vortices, whose motion induces dissipation that adversely affects superconducting properties.  Most real materials have heterogeneous defect structures, a preferential feature\cite{Kwok2016} to produce high critical current densities $J_c$ because no single defect type is effective at pinning vortices over a wide range of temperatures and magnetic fields.  Understanding vortex matter in heterogeneous microstructures poses a tremendous challenge because the effects of combinations of defects are not simply additive, but are often competitive \cite{Kwok2016, Miura2011b}.  Considering systems with only one  defect type, a flux line or bundle of lines can be pinned by the collective action of many weak defects (weak collective pinning)\cite{Larkin1979, Blatter1994b} or the independent action of stronger defects\cite{Larkin1979, Ovchinnikov1991, Blatter2004a}.  In the former case, individual defects are too weak to pin a vortex, while density fluctuations by ensembles of defects contained within a correlation volume $V_c \sim R_c^2 L_c$ (depicted in Fig. \ref{fig:cartoon}) produce a finite pinning force. The volume $V_c$ determines the characteristic pinning energy scale $U_{c}$.  Because collective pinning theory considers only point defects, it is unclear---and beyond the scope of this theory---how $V_c$ and $U_{c}$ are affected by even a sparse distribution of larger defects coexisting with point defects.

A further complication to understanding vortex dynamics arises when thermal fluctuations sufficiently energize vortices to overcome the current-dependent activation energy $U(J)$.  Vortices then depin (creep) from defects causing the persistent current density to decay logarithmically over time $t$ as $J(t) \propto \left[1+ (\mu T/U_c) \ln (t/t_0) \right]^{-1/\mu}$, where $1/t_0$ is a microscopic attempt frequency and $\mu>0$ is the glassy exponent \cite{Malozemoff1990a}. This decay defines a creep rate $S \equiv -d \ln J/ d \ln t = T/[U_c+\mu T \ln(t/t_0)]$. With few exceptions, most studies of vortex creep have been performed on single crystals or thick films (e.g., coated conductors with $d \gtrsim 1 \mu \textnormal{m}$).\cite{Eley2017} Yet the broad, often detrimental impact of vortex motion on thin-film-based devices\cite{Stan2004, Song2009, Wang2014} warrants a better understanding of creep in thin films. For sufficiently thin samples, $d < L_{c}$, the activation energy and creep rate should acquire an explicit dependence on thickness\cite{Larkin1979, Brandt1986b, Kes1987} and the vortex (or bundle) should no longer behave elastically, but rather rigidly. This should produce faster creep. 

In this Letter, we present a systematic study of the dependence of the creep rate on film thickness in two very different superconductors: niobium (Nb) and (Y,Gd)Ba$_2$Cu$_3$O$_{7-x}$. Nb is commonly used for device applications, and features a low critical-temperature $T_c = \textnormal{9.2 K}$,  moderately large Ginzburg-Landau parameter $\kappa=\lambda / \xi \sim 11$ (our film), and low Ginzburg number\cite{Eley2017} $Gi \sim 10^{-8}$.  Here, $\xi$ and $\lambda$ are the in-plane coherence length and penetration depth, respectively. Our film was deposited using DC magnetron sputtering to a thickness of 447 nm, is polycrystalline, and the microstructure includes point defects and grain boundaries. The cuprate (Y,Gd)Ba$_2$Cu$_3$O$_{7-x}$ has high $T_c = \textnormal{92 K}$, $\kappa \sim 95$, and $Gi \sim 10^{-2}$ and is a primary choice for high-current applications. \cite{Wesche2013}  It was grown epitaxially to a thickness of 900 nm using metal organic deposition, and its microstructure consists of point defects, a sparse distribution of Y$_2$Cu$_2$O$_5$ (225) precipitates (diameter $\sim$94 nm, spacing $\sim$272 nm) and twin boundaries, all common in as-grown cuprate films. \cite{Foltyn2007a, Miura2016}

\section{Results and Discussion}

\begin{figure}[t!]
\centering
\includegraphics[scale=0.65]{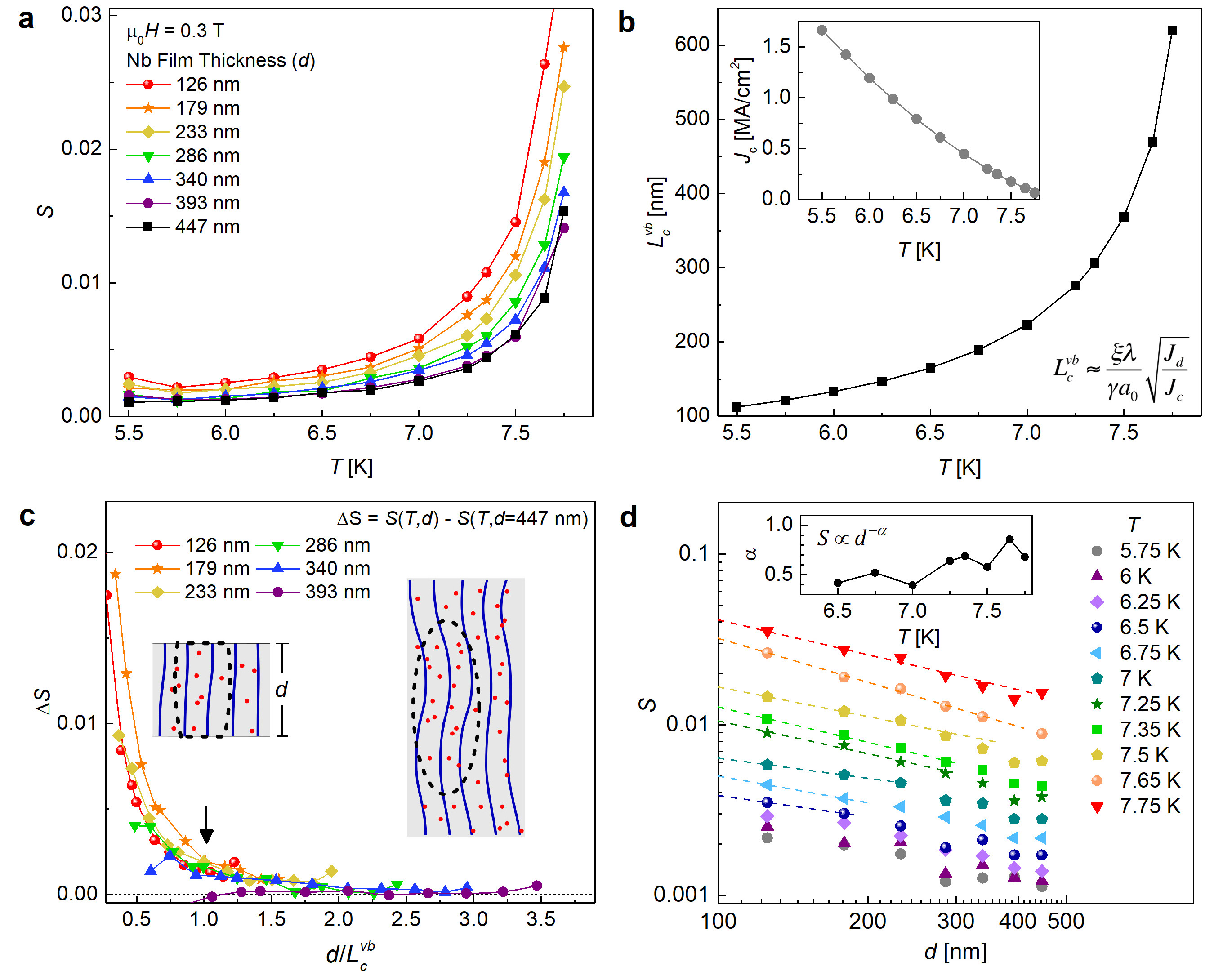}
\caption{\textbf{Thickness-dependent vortex creep in Nb film.} \textbf{a}, Comparison of the temperature-dependent creep rates, $S(T)$, for a Nb film at different thicknesses $d$ in a field of 0.3 T. The error bars [standard deviation in the logarithmic fit to $M(t)$] are smaller than the point sizes. We were unable to measure creep below 5.5 K due to flux jumps. \textbf{b}, Temperature dependence of the collective pinning length for vortex bundles in our Nb film based on equation \eqref{eq:Lc}, our experimental data for $J_c$ (inset), and the assumption that $\lambda(T=0)\approx \textnormal{80 nm}$.\cite{Grebinnik1990a} \textbf{c}, Difference in creep rate of thinned film and that of the original film ($d=\textnormal{447 nm}$) versus the ratio of the film thickness to the collective pinning length, $d / L_c(T)$.  Creep rate deviates from bulk behavior when film becomes thinner than $L_c^{vb}(T)$. The insets illustrate the bulk (right panel) and truncated (left panel) collective pinning volumes, applicable for $d>L_c$ and $d<L_c$, respectively. \textbf{d}, Creep versus film thickness at different temperatures on a $\log-\log$ plot.  The dotted lines are linear fits for $d \leq L_c^{vb}$ used to extract the exponent $\alpha$ (slopes of fits) of the power-law $S \propto d^{-\alpha}$, plotted in the inset.}
\label{fig:figNb}
\end{figure}

Using a SQUID magnetometer, we performed magnetization measurements $M(t) \propto J(t)$ to determine the temperature-dependent creep rate $S(T)$, $J_c$, and $T_c$. We alternated between these measurements and thinning the films using a broad beam Ar+ ion mill to characterize changes in these parameters with decreasing sample thickness.  Further details of the experimental procedures are provided in the Methods section. Figure \ref{fig:figNb}a compares $S(T)$ at different film thicknesses for Nb measured in a $0.3~\mathrm{T}$ field. Creep is similar for $d$ = 447 nm and 393 nm, while subsequent thinning causes successive increases in $S(T)$ that are more pronounced at higher temperatures. This progression from a thickness-independent to thickness-dependent $S(T)$ is suggestive of the sought-after 3D-to-2D transition.  

We now compare the experimental data with predictions from weak collective pinning theory\cite{Larkin1979, Blatter1994b}, which distinguishes two main (bulk) regimes: the pinning volume may contain single vortices (sv) or vortex bundles (vb) with characteristic longitudinal and transverse correlation lengths
\begin{align}\label{eq:Lc}
   L_c &=
   \left\{
   \begin{aligned}
      L_c^{sv} &\approx (\xi/\gamma) (J_d/J_c)^{1/2}\\
      L_c^{vb} &\approx (\xi \lambda / \gamma a_{0}) (J_d/J_c)^{1/2}
   \end{aligned}
   \right.
   &
   R_c &=
   \left\{
   \begin{aligned}
      R_c^{sv} &\approx \xi\\
      R_c^{vb} &\approx \xi (J_d/J_c)^{1/2}
   \end{aligned}
   \right.
   &
   \begin{aligned}
      &\text{for } L_{c}^{sv} < a_{0}\\
      &\text{for } \lambda^{2} / a_{0}  < L_c^{vb}.
   \end{aligned}
\end{align}
Here, $J_d$ is the depairing current density, $a_{0} = (4/3)^{1/4}(\Phi_{0} / B)^{1/2}$ is the vortex spacing, $B$ is the magnetic induction, $\xi$ is the coherence length in the ab-plane, and $\gamma > 1$ is the uniaxial anisotropy parameter. Note that equation \eqref{eq:Lc} neglects the narrow regime of small vortex bundles for which $a_{0} < L_{c} < \lambda^{2}/a_{0}$ and $R_{c}$ rapidly grows from $\xi$ to $\lambda$. The pinning energy governing the creep rate is $U_{c} \approx c_{66}(\xi/R_{c})^{2} V_{c}$, where $c_{66}$ is the shear modulus. Given an applied field of 0.3 T ($a_0 \approx \textnormal{89 nm} \sim \lambda$) and $L_{c}$ determined from critical current density data, the system clearly is in the vortex bundle regime, $L_c^{vb}(T) > \lambda^2/a_{0}$, see Fig. \ref{fig:figNb}b. Based on these findings, we re-plot the data from Fig. \ref{fig:figNb}a to display the change in creep rate from that of the original film ($\Delta S$) versus $d/L_{c}^{vb}(T)$, see Fig. \ref{fig:figNb}c.  We observe that $S$ abruptly deviates from bulk behavior when the film becomes thinner than $L_{c}(T)$, revealing the 3D-to-2D transition.  

Two-dimensional collective pinning was first experimentally realized in studies\cite{Kes1983, Wordenweber1986, Toyota1984, Yoshizumi1984, Osquiguil1985} of the field-dependent pinning force in amorphous superconducting films of thickness $d \ll L_c$. 
On the 2D side of the transition, the longitudinal correlation length is capped by the sample thickness, $L_{c} \to d$, and the transverse correlation length\cite{Brandt1986b, Kes1987}
\begin{numcases}{R_c=}
R_c^{sv} \approx \xi & for $d < a_0$, \label{eq:Rc2Dsv} \\
R_c^{vb} \propto d^{1/2}  & for $a_{0} < d $,\label{eq:Rc2Dlb}
\end{numcases}
replace the bulk expressions in Eq.~\eqref{eq:Lc}. The energy $U_{c}$ then scales as
\begin{numcases}{U_c \propto }
U_c^{sv} \propto d^{1/2} & for $d < a_0$, \label{eq:Uc2Dsv} \\
U_c^{vb} \propto d & for $a_{0} < d $.\label{eq:Uc2Dlb}
\end{numcases}
Plotting $S(d)$ for fixed temperatures, Fig. \ref{fig:figNb}d, we observe a relatively flat ($d$-independent) region at nearly all temperatures for large values of $d$ and power law behavior $S \propto d^{-\alpha}$ for small values of $d$. Using $S \sim T/U_{c}$, the exponent $0.5 < \alpha < 1$ (Fig. \ref{fig:figNb}d inset) assumes values in agreement with equations\ \eqref{eq:Uc2Dsv} and \eqref{eq:Uc2Dlb}.

The creep data further allows us to define the effective activation energy $U^{*} \equiv T/S = U_{c} + \mu T \ln (t/t_0)$, hence providing direct access to the glassy exponent $\mu$. This exponent captures the diverging behavior of the  current-dependent activation barrier $U(J) \sim U_{c} (J_{c}/J)^{\mu}$ away from criticality, i.e., for $J \ll J_{c}$, and assumes different values for creep of single vortices versus bundles\cite{Blatter1994b}. Near criticality, the behavior is non-glassy and the exponent is named $p < 0$ instead of $\mu$. A detailed analysis of the system's glassiness is provided in the Supplementary Information (see Fig. S5) and further confirms the transition from elastic to rigid vortex behavior at $d \approx L_{c}$.

\begin{figure}[t!]
\centering
\includegraphics[scale=0.55]{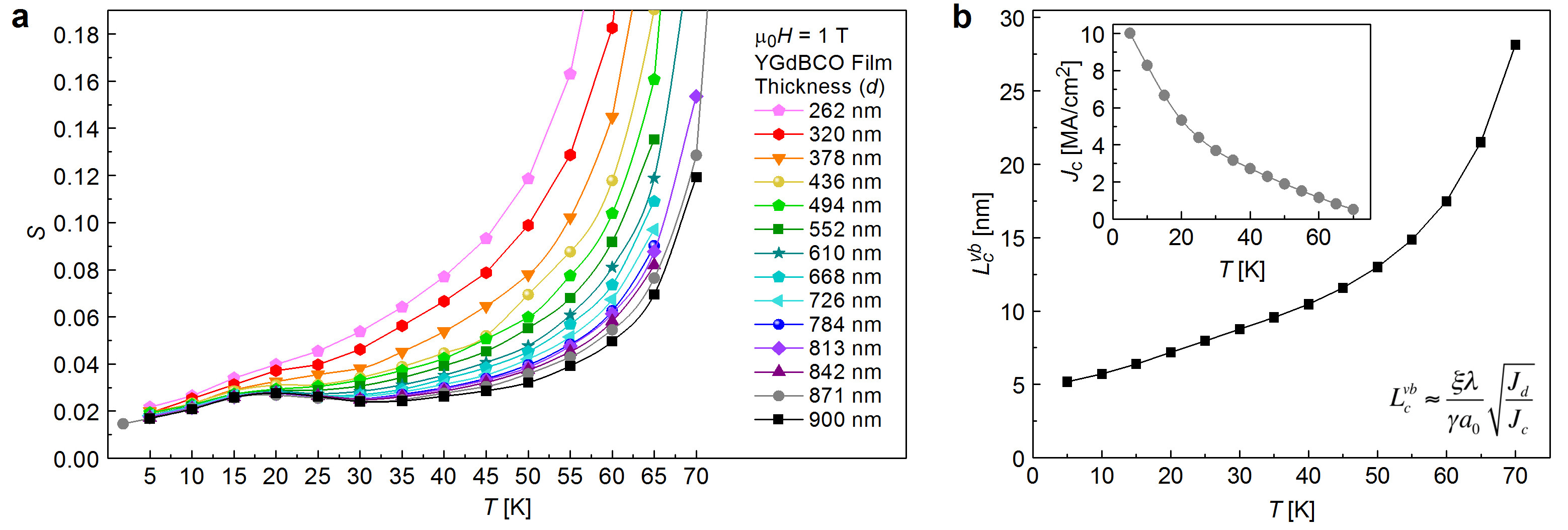}
\caption{\textbf{Thickness-dependent vortex creep in (Y,Gd)Ba$_2$Cu$_3$O$_{7-x}$ film.} \textbf{a}, Comparison of the temperature-dependent creep rates, $S(T)$, for the (Y,Gd)Ba$_2$Cu$_3$O$_{7-x}$ (YGdBCO) film at different thicknesses $d$ in a field of 1 T. The error bars [standard deviation in the logarithmic fit to $M(t)$] are smaller than the point sizes. \textbf{b},  Collective pinning length $L_c^{vb}(T)$ for (Y,Gd)Ba$_2$Cu$_3$O$_{7-x}$ in the bundle regime based on equation \eqref{eq:Lc}, our experimental data for $J_c$ (inset), and the assumption that $\lambda_{ab}(0) \approx \textnormal{140 nm}$ is equivalent to YBa$_2$Cu$_3$O$_{7-x}$ \cite{Farber1998, Langley1991}.  The calculated pinning length is too small to account for the dependence $S(d)$ observed in \textbf{a}.}
\label{fig:figYGdBCOSvsd}
\end{figure}

We performed a complementary study on the cuprate film at 1 T. The evolution of $S(T)$ with thinning, plotted in Fig.\ \ref{fig:figYGdBCOSvsd}a, again reveals successive increases in $S$ that are more pronounced at higher temperatures.  In the original film, $S(T)$ is non-monotonic, showing a shallow dip between 25 K and 45 K; this is typical of YBa$_2$Cu$_3$O$_{7-x}$ and has been associated with the presence of large precipitates.\cite{Haberkorn2012d}  Notwithstanding a fixed density of precipitates at all thicknesses, the dip disappears with thinning, indicative of changes in vortex-precipitate interactions. In contrast, when the film is thinnest ($d$=262 nm) the creep rate grows monotonically with $T$ and qualitatively adheres to the Anderson-Kim model describing creep of rigid vortices.\cite{Anderson1964}

The calculated length $L_c^{vb}(T)$, shown in Fig.\ \ref{fig:figYGdBCOSvsd}b, is far too small to reconcile the observed thickness dependence of $S$. We hence conclude that, though sparse, the large 225 precipitates of size $b \gg L_{c}^{vb}$ must substantially influence effective pinning scales.\cite{Gurevich2007, Koshelev2011a, Willa2017a}  Accounting for these inclusions as strong defects, the pinning length cannot be smaller than $b$. This raises an important question: can a meaningful pinning length $L_{c}^{mp}(T)$ (mixed pinning, mp) be extracted from the experimental data in heterogeneously disordered samples?

\begin{figure}[t!]
\centering
\includegraphics[width=0.9\textwidth]{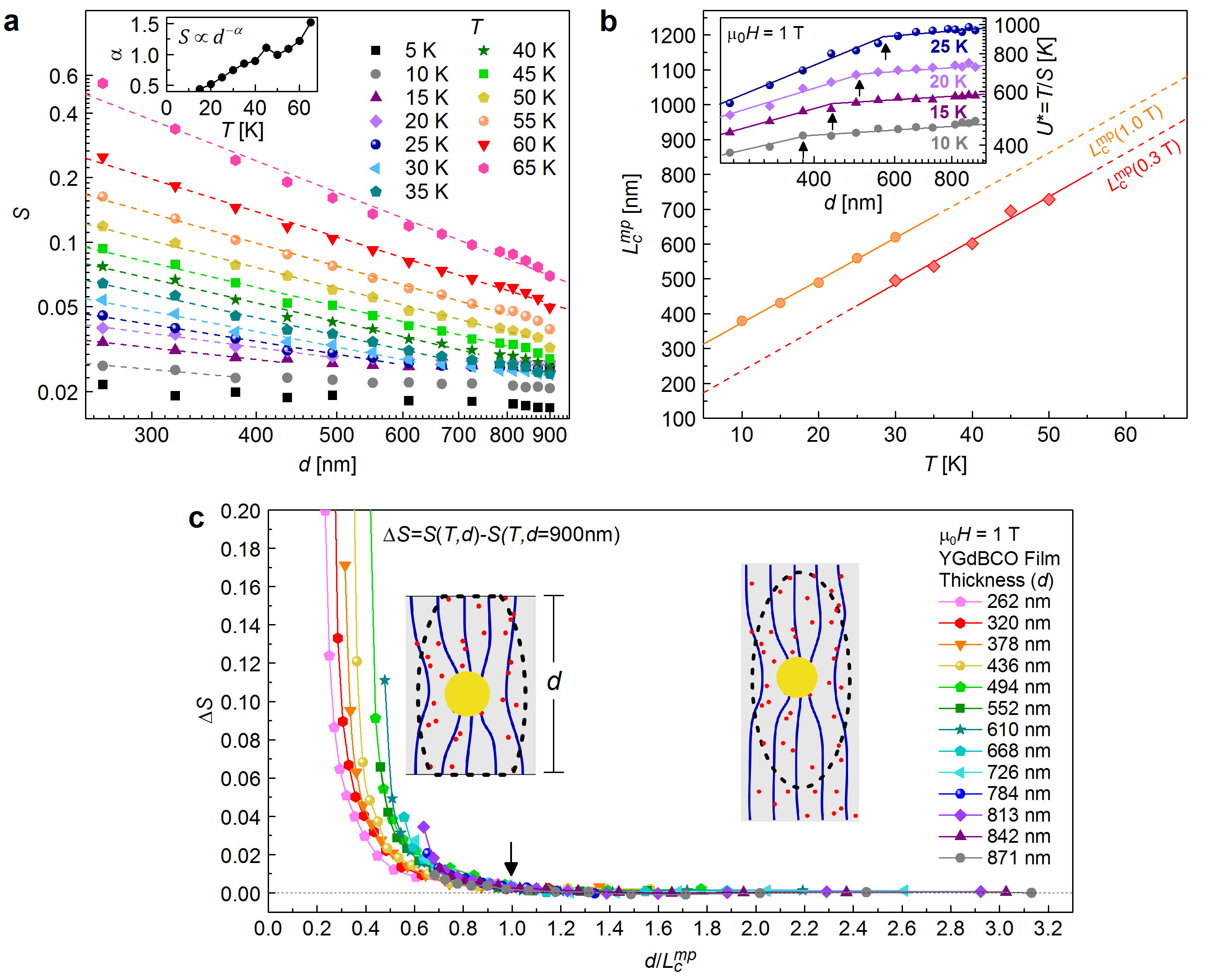}
\caption{\textbf{Pinning length in (Y,Gd)Ba$_2$Cu$_3$O$_{7-x}$ film.} \textbf{a}, Creep rate versus film thickness at different temperatures for (Y,Gd)Ba$_2$Cu$_3$O$_{7-x}$ (YGdBCO) film on a $\log-\log$ plot.  The dotted lines are linear fits for $d \leq L_c^{mp}(T)$.  \textbf{b}, Effective pinning length $L_c^{mp}$ at 0.3 T and 1 T. (See Supplemental Information for all 0.3 T data).  The circles and diamonds show points extracted from the $U^{*}(d)$ data. The \textbf{inset} shows an example of how $L_c^{mp}$ was extracted for $T$ = 10-25 K at 1 T.  Linear fits to this data (solid lines) and extrapolations (dashed lines) defining $L_c^{mp}(T, H)$ are shown. \textbf{c}, Difference in creep rate of thinned film and that of the original film ($d =\textnormal{900 nm}$) versus the ratio of the film thickness to the effective pinning length, $d / L_c^{mp}(T)$.  Creep rate increases sharply when film becomes thinner than $L_c^{mp}$. The insets illustrate the bulk (right panel) and truncated (left panel) pinning volumes, applicable for $d>L_c^{mp}$ and $d<L_c^{mp}$, respectively. 
}
\label{fig:figYGdBCOLb}
\end{figure}

Again, plotting $S(d)$ on a $\log-\log$ scale (Fig.\ \ref{fig:figYGdBCOLb}a), we observe that at low temperatures $S$ is relatively independent of $d$ and conclude that creep is bulk-like, i.e. $L_{c} < d$ for all thicknesses. At higher temperatures, $S$ and therefore the energy $U^{*}$ acquire a thickness dependence, which follows the power law $U^{*} \propto d^\alpha$.  The transition from the thickness-independent to thickness-dependent $U^{*}$ appears as a kink in $U^{*}(d)$, as shown in the inset to Fig.\ \ref{fig:figYGdBCOLb}b.  We identify $L_{c}^{mp}$ as the position of this kink and find a linear dependence on temperature (Fig.\ \ref{fig:figYGdBCOLb}b). This length decreases with decreasing magnetic field, as concluded from a similar analysis of data collected at 0.3 T, presented in detail in the Supplementary Information.  This trend agrees with theoretical predictions within strong pinning theory\cite{Willa2017a}.

Figure \ref{fig:figYGdBCOLb}c highlights the effectiveness of our method: there is an abrupt departure in $S$ from that of the original film when the thickness falls below the extracted pinning length. We conclude that in the prevailing case of heterogeneously disordered superconductors, for which collective pinning by small defects and strong pinning by large precipitates coexist, a pinning length is indeed well defined. This length is much larger than predicted by collective pinning theory, consistent with theoretical expectations. \cite{Gurevich2007, Koshelev2011a, Willa2017a} 

\begin{figure}[t!]
\centering
\includegraphics[width=0.9\textwidth]{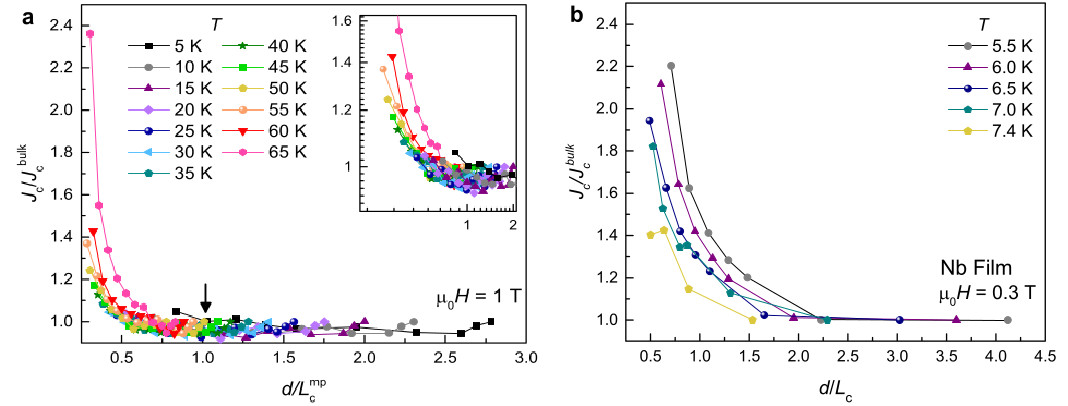}
\caption{\textbf{Increase in critical current with decreasing film thickness.} \textbf{a}, $J_c / J_c^{bulk}$ at different temperatures versus the ratio of the film thickness to the effective pinning length, $d / L_c^{mp}(T)$.  Here, $J_c$ is extracted using a criterion of $t_i = 5 $s (see Supplementary Information for details) and $J_c^{bulk} \equiv J_c( d = 871 \textnormal{ nm})$ at each specified temperature.  $J_c$ increases sharply when film becomes thinner than $L_c^{mp}$. \textbf{b}, $J_c / J_c^{bulk}$ for a Nb film cut from the same substrate as the film in which creep was measured.  Here $J_c^{bulk} \equiv J_c( d = 500 \textnormal{ nm})$ and the data displayed represents film thicknesses $d =$~500 nm, 270 nm, 164 nm, 138 nm, 111 nm, 85 nm, and 58 nm.}
\label{fig:figYGdBCOJcd}
\end{figure}

The magnetic 3D-to-2D crossover should invoke a change in $J_c$ versus $d$. Figure  \ref{fig:figYGdBCOJcd}a shows the normalized $J_c$ versus $d/L_c^{mp}$, using a criterion of $J_c \equiv J(t_i = 5 s)$ (see Supplementary Information for details) and where $L_c^{mp}$ is extracted from the creep measurements as discussed above. Around $d/L_c^{mp} \sim 1$, there is an abrupt transition from a thickness-independent to a thickness dependent regime, at which point $J_c$ increases with decreasing $d$. We observed similar behavior in a Nb film around $d/L_c$, for which results are displayed in Fig. \ref{fig:figYGdBCOJcd}b.


Governing the energy barrier to vortex motion, the pinning length plays a major role in the emergence of fascinating phenomena such as glassiness and plasticity. It is therefore a crucial parameter to consider when pursuing the ambitious materials-by-design paradigm and development of a universal description of vortex matter. This work opens the door for further studies to unveil pinning lengths in other superconductors and could impact work on other problems that similarly involve the interplay between disorder and collective interactions, such as systems containing skyrmions \cite{Reichhardt2015}, domain walls \cite{Lemerle1998}, or disordered polymers. \cite{Halpin-Healy1995}

\vskip 2em
\section*{Methods}
\noindent \textbf{Film Growth and Characterization.} The (Y$_{0.77},$Gd$_{0.23}$)Ba$_2$Cu$_3$O$_{7-x}$ film measured in this study was grown epitaxially on buffered Hastelloy substrates using metal organic deposition from Y-, Gd-, and Ba-trifluoroacetates and Cu-naphthenate solutions.  A stack of NiCrFeO, Gd$_2$Zr$_2$O$_7$, Y$_2$O$_3$, MgO (deposited using ion beam assisted deposition), LaMnO$_3$, and CeO$_2$ layers form the interposing buffer.  Characterized by transmission electron microscopy (TEM) and energy dispersive spectroscopy, the film contain a sparse distribution (3$\times 10^{19}$/m$^3$) of Y$_2$Cu$_2$O$_5$ precipitates, which are on average 94 nm in diameter and spaced 272 nm apart (see ref. [\citen{Miura2016}] for TEM images).  The Nb film was deposited on a SiO$_2$ substrate by DC sputtering and contains $\sim$10 nm sized grains \cite{Henry2014}.  As discussed in the main text, Nb has low $\kappa$ and low $Gi$, while (Y$_{0.77},$Gd$_{0.23}$)Ba$_2$Cu$_3$O$_{7-x}$ has both high $\kappa$ and $Gi$.  The relevance of these parameters is as follows: $\kappa$ is related to the energy required for vortex core formation [almost negligible in (Y,Gd)Ba$_2$Cu$_3$O$_{7-x}$] and $Gi$ captures the relevance of thermal fluctuations in the material, i.e., materials with larger $Gi$ suffer more from vortex creep \cite{Ginzburg1961, Blatter1994b, Eley2017}.   

Thinning was achieved by broad beam Ar$^+$ ion milling on a rotating, water-cooled stage.  To prevent significant changes in the surface roughness, we milled at a low beam voltage of 300 V and current of 12 mA.  Atomic force microscopy studies revealed milling rates of $\sim \textnormal{12 nm} / \textnormal{min}$ and $\sim \textnormal{9 nm} / \textnormal{min}$ and RMS surface roughnesses of 30 nm and 5 nm for the (Y,Gd)Ba$_2$Cu$_3$O$_{7-x}$ film and Nb film, respectively.   As a redundant verification of the rate calibration, the thickness of the (Y,Gd)Ba$_2$Cu$_3$O$_{7-x}$ film after all milling steps was measured using focused ion beam and cross sectional scanning electron microscopy.
\vskip 1em

\paragraph{\textbf{Magnetization Measurements.}} Magnetization studies were performed using a Quantum Design SQUID magnetometer.  For all measurements, the magnetic field was applied perpendicular to the film plane.  Creep data were taken using an established approach\cite{Yeshurun1996b} where the decay in the magnetization over time was captured by repeatedly measuring $M$ every 15 s at a fixed $T$ and $H$.  To accomplish this, the field was initally swept high enough ($\Delta H > 4H^*$) that vortices, which first form at the film peripheries, permeate the center of the sample. In this case, $H^*$ is the field of full flux penetration and the Bean critical state model defines the vortex distribution.  A brief measurement of $M(t)$ in the lower branch was collected, then for the upper branch $M$ was repeatedly measured for an hour.  After subtracting the background (determined by the differences in the upper and lower branches) and adjusting the time to account for the difference between the initial application of the field and the first measurement (maximize correlation coefficient), $S=-\dd \ln M / \dd \ln t$ was calculated from the slope of a linear fit to $\ln M $ versus $\ln t$.  The critical current density was calculated from the magnetization data using the Bean model \cite{Bean1964a, Gyorgy1989a}, $J_c=20 \Delta M / [w(1-w/3l)]$. Here, $\Delta M$ is the difference between the first few points from the upper and lower branches of $M(t)$ , $w \sim l \sim 4-5$ mm specifies sample width and length. $T_c$ was determined from temperature-dependent magnetization curves recorded at 0.002 T. For both the Nb film and the (Y,Gd)Ba$_2$Cu$_3$O$_{7-x}$ film, thinning did not change $T_c$.  
\vskip 1em

\section*{Acknowledgments} All measurements (S.E.), work performed by S.E. at the Center for Integrated Nanotechnologies (CINT), contributions from L.C., and theory (R.W.) were funded by the U.S. Department of Energy (DOE), Office of Basic Energy Sciences, Materials Science and Engineering Division.  CINT is a user facility operated for the DOE Office of Science.  M.M. and M.S. are supported by JSPS KAKENHI (17H03239 and 17K18888) and a research grant from the Japan Power Academy. M.D.H. was supported by the LDRD program at Sandia National Laboratory (SNL). The authors acknowledge and thank the staff of the SNL MESA fabrication facility.  We also thank Dr.\ Charles Thomas Harris of CINT at SNL for focused ion beam milling and cross-sectional scanning electron microscopy. R.W. also acknowledges support from the Swiss National Science Foundation (SNSF) through the Early Postdoc.Mobility program.

\vskip 1em

\section*{Author Information}

\subsection*{Author Contributions} S.E. conceived and designed the experiment, performed the magnetization measurements, ion milling, atomic force microscopy, data analysis, and assisted with theoretical analysis. R.W. spearheaded the theoretical analysis. S.E. and R.W. wrote the manuscript. L.C. assisted with theoretical analysis and manuscript preparation. M.L. assisted with data collection. M.M. and M.S. grew and performed the microstructure characterization on the (Y,Gd)Ba$_2$Cu$_3$O$_{7-x}$ films. M.D.H. deposited and performed the microstructure characterization on the Nb films. 
\vskip 1em

\subsection*{Competing Financial Interest Statement} 
The authors declare no competing financial or non-financial interests.

\subsection*{Data Availability} 
Data supporting the findings in this study are available within the article or from the corresponding author upon reasonable request.

\pagebreak

\bibliographystyle{naturemag}

\vskip 1em

\newpage

\end{document}